\documentclass[aps, prb, twocolumn,groupedaddress]{revtex4} 
\usepackage{graphicx}
\usepackage{amsmath}
\usepackage{amssymb}
\usepackage{ifthen}
\usepackage{booktabs}

\newcommand{\bb}{\begin{equation}}
\newcommand{\ee}{\end{equation}}
\newcommand{\ba}{\begin{eqnarray*}}
\newcommand{\ea}{\end{eqnarray*}}
\newcommand{\rhor}{\rho({\bf r})}
\newcommand{\dd}{{\rm d}}
\newcommand{\rr}{{\mathbf r}}

\usepackage{graphicx}
\usepackage{dcolumn}
\usepackage{bm}
\usepackage{longtable}

\begin{document}


\title{Density functional study of complete, first-order and critical wedge filling transitions}

\author{Alexandr \surname{Malijevsk\'y}}
\affiliation{
{E. H{\'a}la Laboratory of Thermodynamics, Institute of Chemical Process Fundamentals, Academy of Sciences, 16502 Prague 6, Czech Republic}\\
{Department of Physical Chemistry, Institute of
Chemical Technology, Prague, 166 28 Praha 6, Czech Republic}}
\author{Andrew O \surname{Parry}}
\affiliation{Department of Mathematics, Imperial College London, London SW7 2B7, UK}

\begin{abstract}

We present numerical studies of complete, first-order and critical wedge filling transitions, at a right angle corner, using a microscopic fundamental measure
density functional theory. We consider systems with short-ranged, cut-off Lennard-Jones, fluid-fluid forces and two types of wall-fluid potential: a purely
repulsive hard wall and also a long-ranged potential with three different strengths. For each of these systems we first determine the wetting properties
occurring at a planar wall including any wetting transition and the dependence of the contact angle on temperature. The hard wall corner is completely filled by
vapour on approaching bulk coexistence and the numerical results for the growth of the meniscus thickness are in excellent agreement with effective Hamiltonian
predictions for the critical exponents and amplitudes, at leading and next-to-leading order. In the presence of the attractive wall-fluid interaction, the
corresponding planar wall-fluid interface exhibits a first-order wetting transition for each of the interaction strengths considered. In the right angle wedge
geometry the two strongest interactions produce first-order filling transitions while for the weakest interaction strength, for which wetting and filling occur
closest to the bulk critical point, the filling transition is second-order. For this continuous transition the critical exponent describing the divergence of the
meniscus thickness is found to be in good agreement with effective Hamiltonian predictions.

\end{abstract}


\maketitle

\section{Introduction}
Wetting transitions and related fluid interfacial phenomena have been extensively studied over the last few decades (see for example the excellent review articles \cite{dietrich, sullivan, schick, bonn, saam}). The vast
majority of early theoretical studies focussed on fluid adsorption on idealized planar substrates, or between parallel plates \cite{evans86, evans90} or around spheres and cylinders \cite{indekeu, stewart, parry06, nold}, in
which the equilibrium density profile is one dimensional and depends only on the coordinate normal to the substrate. More recently however there has been considerable interest in adsorption and wetting at micro-patterned
surfaces in which the substrate is non-planar \cite{rascon_nature} or is chemically heterogeneous \cite{bauer}. This work has been motivated mainly by improvements in surface lithography and related techniques which now allow
the controlled fabrication of tailored substrates which are central to the development of microfluidics. At a more fundamental level however, such studies have revealed new examples of interfacial phase transitions and
fluctuation effects, as well as surprising connections between adsorption in different geometries \cite{abraham, parry03, rascon05, morgan, tas, tas2, rascon10, parry11}.

\begin{figure}[ht]
\includegraphics[width=8.5cm]{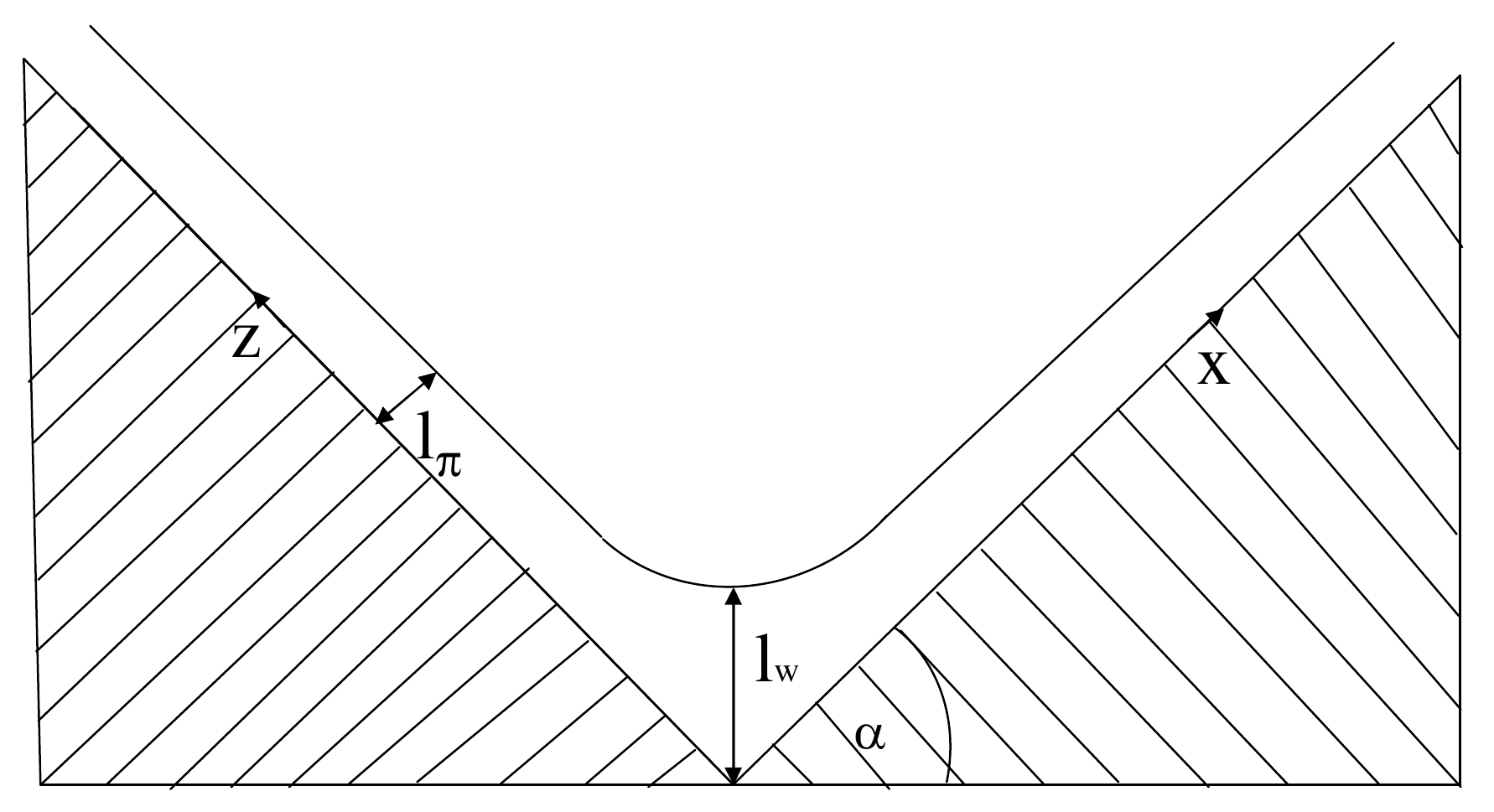}
\caption{Schematic picture of the cross section of a right angle wedge for which $\alpha=\pi/4$. Far from the apex, the wetting layer is of thickness $\ell_\pi$.
However, the height of the meniscus above the apex is $\ell_w$. At a filling transition $\ell_w$ changes from microscopic to macroscopic. The Cartesian
coordinates $x$ and $z$ used in our analysis are shown. }\label{wedge_pic}
\end{figure}

A particularly simple and important example of a non-planar substrate is a wedge geometry formed by two identical infinite planar walls that meet at an opening angle $ 2\psi=\pi-2\alpha$ where $\alpha$ is the tilt angle with
respect to the horizontal plane (say). In Fig.\,\ref{wedge_pic} we schematically show a section of a three-dimensional wedge where the walls meet at a right angle corresponding to $\alpha=\pi/4$. The wedge geometry may be
thought as being a missing link between the very well studied examples of a planar wall ($\alpha=0$) and a capillary-slit ($\alpha=\pi/2$) and shows a phase transition which is distinct from wetting and capillary condensation.
Let us suppose that the substrate is in contact with a bulk vapour at chemical potential $\mu$, tuned to saturation $\mu=\mu_{\rm sat}^-$, and at a temperature $T$ less than the bulk critical temperature $T_c$. Gravity is
ignored. Macroscopic arguments, which have been discovered independently by several authors \cite{shuttle, concus, pomeau, hauge}, dictate that the wedge is completely filled with liquid when $\theta<\alpha$ where $\theta(T)$
is the contact angle defined for a macroscopic sessile drop on a flat surface. However for $\theta>\alpha$ the adsorption of liquid at the wedge is microscopic. The {\it{wedge filling transition}} corresponds to the transition
from microscopic to macroscopic preferential adsorption of liquid, at a filling temperature $T_f$, which satisfies the exact condition
 \bb
 \theta(T_f)=\alpha\,. \label{fill}
 \ee

Since the contact angle usually decreases with temperature it follows that $T_f<T_w$ where $T_w$ is the wetting temperature at which the contact angle vanishes.
In other words wedge filling precedes wetting i.e. the wedge can be completely filled with liquid even though the walls are only partially wet. In Fig.\,2 we
show two possible phase diagrams illustrating first-order and continuous wedge filling transitions. In each case the filling transition refers to the change from
microscopic to macroscopic adsorption as $T\to T_f$ along the coexistence line $\mu=\mu_{\rm sat}^-$. In Fig.\,2a we suppose this transition is first-order while
in Fig.\,2b we suppose it is continuous (critical filling). In the latter case the equilibrium height $\ell_w$ of the meniscus above the wedge bottom diverges
continuously in this limit. For the case of first-order filling, a pre-filling line (shown as dotted), corresponding a thin-thick transition extends above $T_f$
and off coexistence, analogous to the pre-wetting line which is also shown. However, unlike pre-wetting, the pre-filling transition is necessarily rounded since
it is pseudo one dimensional and thus the pre-filling line does not end in a genuine critical point. Both phase diagrams also show the {\it{complete filling}}
transition which corresponds to the continuous divergence of the meniscus height as $\mu\to\mu_{\rm sat}^-$ for $T>T_f$.

\begin{figure}[ht]
\includegraphics[width=8.5cm]{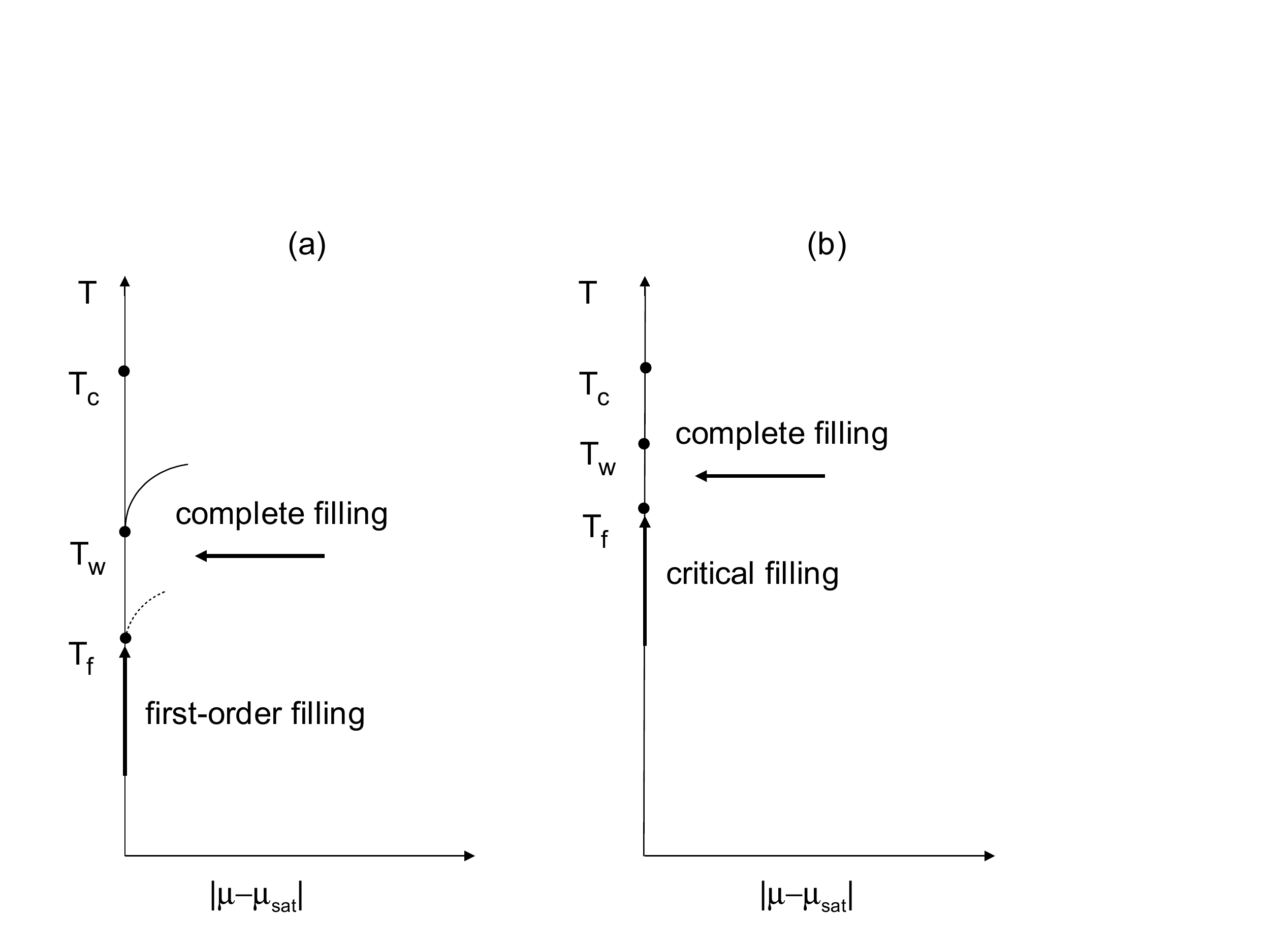}
\caption{Schematic phase diagrams for wetting and filling at a wedge-vapour interface; (a) first-order wetting and filling transitions, (b) continuous wetting
and filling transitions. If the wetting transition is weakly first-order the filling transition may be continuous (critical) in which case the pre-filling line
(dotted) is absent. }\label{filling_scheme}
\end{figure}

Over the last decade, effective interfacial Hamiltonian models have been used extensively to study the order of wedge filling transitions and have shown how these are sensitive to the range of the intermolecular forces and
also interfacial fluctuation effects \cite{rejmer, wood1, wood2, wood3, wood4, abraham02, abraham03, romero05, parry05, romero07, bernardino, rascon11}. These turn out to be rather subtle issues. For example while in open
wedges (small $\alpha$) the order of the transition is qualitatively the same as that of the underlying wetting transition, the critical exponents which characterize three dimensional critical filling are quite different to
those of critical wetting and fluctuation effects are much larger. For acute wedges on the other hand effective Hamiltonians predict that the order of the filling transition may be different to that of the wetting transition
\cite{wood2, bernardino}. While some of these predictions have been verified in computer simulations \cite{binder03, binder05}, exact Ising model calculations \cite{abraham02, abraham03} and simple square-gradient mean-field
theory \cite{bernardino}, to the best of our knowledge the filling transition has not been studied using modern microscopic classical density functional theory (DFT) \cite{evans_79} or for systems with realistic long-ranged
intermolecular forces. DFT has proved an invaluable tool in the study of inhomogeneous fluids including interfacial properties, wetting, layering and capillary condensation transitions. As mentioned above, studies of such
transitions are simplified because the density profile is one dimensional. The purpose of this paper is to apply modern DFT to the study of the wedge filling transition for which of course the density profile is two
dimensional. In this way our work complements recent studies of condensation in capped capillaries \cite{roth, mal}. A preliminary account of some of our results has appeared earlier \cite{our_prl}.

In our work we use Rosenfeld's fundamental measure theory (FMT) \cite{ros, roth_fmt} combined with a simple mean-field treatment of the attractive part of the intermolecular interaction, which is taken to be a cut-off
Lennard-Jones potential. We consider a right-angle wedge and two different types of wall-fluid interaction which allow us to address three specific points and compare with effective Hamiltonian theory. 1) For a purely hard
wall, for which the contact angle $\theta=\pi$, we study predictions for universal critical singularities for the complete wedge filling (of gas). In this case the complete filling occurs at the wedge-liquid interface as
$\mu\to\mu_{\rm sat}^+$. 2) We consider walls with long-ranged wall-fluid interactions which leads to a finite contact angle and a wetting transition at which the contact angle vanishes. The wetting transition is always
first-order but the strength of it is weakened the closer the transition occurs to the bulk critical temperature. In this way we can see if the order of the filling transition can change and be different to the order of the
underlying wetting transition. If this is the case we wish to extract the critical exponents and compare with effective Hamiltonian predictions.

Our paper is organized as follows. We start with a description of our model DFT, the intermolecular forces, the wedge geometry and boundary conditions used. We
first consider the case of complete wedge filling occurring at the interface between a hard wall wedge and a bulk liquid. We check that our numerical results
satisfy exact sum-rules for a planar hard wall, using the full 2D code, and then extract the equilibrium meniscus shape and excess adsorption for the right angle
wedge geometry and compare with effective Hamiltonian predictions for critical exponents and critical amplitudes. We then add an attractive long-ranged
wall-fluid potential and first determine the contact angle and wetting transition temperature for a planar wall-gas interface. For the corresponding wedge
geometry we determine numerically the location and order of the filling transition and compare with the thermodynamic and effective Hamiltonian predictions. We
finish with a summary of our results and discuss some open questions.

\section{Density Functional Theory and Model Interactions}
In this section we describe our model and outline the main features of the microscopic DFT that have been used in this work.


Within classical density functional theory \cite{evans_79}, the equilibrium density profile is found by minimizing the grand potential functional
 \bb
 \Omega[\rho]={\cal F}[\rho]+\int\dd\rr\rhor[V(\rr)-\mu]\,,\label{om}
 \ee
where $\mu$ is the chemical potential and $V(\rr)$ is the external potential. Here ${\cal F}[\rho]$ is the intrinsic free energy functional of the fluid one-body
density, $\rhor$, which can be split into ideal and excess parts. Following the spirit of van der Waals, modern DFT often further divides the latter into a
hard-sphere and an attractive contribution
  \bb
  {\cal F}_{\rm ex}[\rho]={\cal F}_{\rm hs}[\rho]+\frac{1}{2}\int\int\dd\rr\dd\rr'\rhor\rho(\rr')u_{\rm a}(|\rr-\rr'|)\,, \label{f}
  \ee
where  $u_{\rm a}(r)$ is the attractive part of the fluid-fluid interaction potential. In our analysis we take this to be a truncated Lennard-Jones-like
potential
 \bb
 u_{\rm a}(r)=\left\{\begin{array}{cc}
 0\,;& r<\sigma\,,\\
-4\varepsilon\left(\frac{\sigma}{r}\right)^6\,;& \sigma<r<r_c\,,\\
0\,;&r>r_c\,.
\end{array}\right.
 \ee
which is cut-off at $r_c=2.5\,\sigma$, where $\sigma$ is the hard-sphere diameter. The hard-sphere part of the excess free energy is approximated by the FMT
functional \cite{ros},
 \bb
{\cal F}_{\rm hs}[\rho]=\frac{1}{\beta}\int\dd\rr\,\Phi(\{n_\alpha\})\,,
 \ee
where $\Phi$ is a function of six weighted densities $n_\alpha(\rr)$, and $\beta=1/k_BT$ is the inverse temperature. Rosenfeld's FMT accurately captures
short-range correlations and thus the functional (\ref{f}) should describe strong packing effects for liquid adsorption at the surface of the wall and near the
apex.

The confining wedge is treated as an external field, $V(\rr)$, exerted on the fluid atoms. The potential is assumed to be translationally invariant along the
wedge which is formed from two semi-infinite planar slabs (walls) that meet at a right angle, so that $\alpha=\pi/4$. We will consider two types of wedge-fluid
interaction. One is a purely hard wall wedge, whose potential is simply
 \bb
 V^{\rm hw}(x,z)=\left\{\begin{array}{cc} \infty\,;&x<\sigma\;{\rm or}\;z<\sigma,\\
0\,;& {\rm elsewhere\,,}\end{array}\right. \label{hard_wedge}
 \ee
 where the $x$ and $z$ Cartesian coordinates run parallel to the left and right hand side walls respectively (see Fig.~1).

The second wall potential is long-ranged and is assumed to arise from a uniform distribution of wall atoms, with a one-body density $\rho_w$. These interact with
the fluid atoms via the Lennard-Jones potential
 \bb
 \phi_w(r)=-4\varepsilon_w\left(\frac{\sigma}{r}\right)^{6}; \hspace{1cm}r>\sigma\,.
 \ee

After integrating $\phi_w(\rr)$ over the whole depth of the wall, the potential of the wedge can be expressed as
 \bb
 V^{\rm LJ}(x,z)=\left\{\begin{array}{cc} \infty\,;&x<\sigma\;{\rm or}\;z<\sigma,\\
\tilde{V}(x,z)\,;& {\rm elsewhere\,,}\end{array}\right. \label{wedge}
 \ee
 with
 \bb
 \tilde{V}(x,z)=\alpha_w\left[\frac{1}{z^3}+\frac{2z^4+x^2z^2+2x^4}{2x^3z^3\sqrt{x^2+z^2}} +\frac{1}{x^3}\right]\label{pot_wedge}
 \ee
 and
 \bb
\alpha_w=-\frac{1}{3}\pi\varepsilon_w\rho_w\sigma^6\,.
 \ee
Notice that infinitely far from the wedge apex, the potential close to either surface recovers that of a planar wall $V(x,\infty)=2\alpha_w/x^3$ or
$V(\infty,z)=2\alpha_w/z^3$. Minimization of (\ref{om}) leads to an Euler-Lagrange equation which is solved numerically. This is done on an $L\times L$ Cartesian
grid where the lateral dimension of our box size is $L=50\sigma$ and the grid has discretization size $0.05\,\sigma$. To mimic the bulk boundary conditions we
impose that $\rho(L,z)=\rho_\pi(z;L)$ and $\rho(x,L)=\rho_\pi(x;L)$ where $\rho_\pi(z;L)$ is the equilibrium profile for a planar wall-fluid interface with
$\rho_\pi(L;L)$ fixed to the bulk density. The latter is, for the sake of numerical consistency, determined from the full 2D DFT. Once the equilibrium density
profile is found, the corresponding grand potential is calculated from (\ref{om}). From this, all the thermodynamical properties of the system can be determined.
For the most part we express our temperature scale in fractions of the bulk critical temperature $k_BT_c/\varepsilon=1.414$ or in dimensionless units
$T^*=k_BT/\varepsilon$ where more convenient. Similarly, densities are written in dimensionless units $\rho^*=\rho\sigma^3$ as are wetting film thicknesses
$\ell^*=\ell/\sigma$ and distances $z^*=z/\sigma$, etc.

\section{Numerical results}

\subsection{Complete Filling at a Hard Wall Wedge}

As described in the introduction, the complete filling transition refers to the continuous divergence of the meniscus height $\ell_w$, on approaching two phase
coexistence when the contact angle $\theta<\alpha$. Effective Hamiltonian studies predict that this transition is dominated by the geometry of the wedge and
displays universal critical properties \cite{rascon_nature}. For example, at leading order the meniscus height is predicted to diverge as
\begin{equation}
\ell_w\approx\frac{\gamma(\sec\alpha\cos\theta-1)}{\delta\mu\Delta\rho} \label{CoFilling1}
\end{equation}
where $\gamma$ is the surface tension of the liquid-gas interface and $\Delta\rho=\rho_l-\rho_g$ is the difference between the bulk densities. The power-law
dependence on $\delta\mu=|\mu-\mu_{\rm sat}|$ is universal and is independent of the range of the intermolecular forces and fluctuation effects. This universal
behaviour can be understood very simply using macroscopic concepts \cite{rascon_nature}. As coexistence is approach the meniscus that grows at the wedge corner
must have a circular cross-section with radius $R=\gamma/\delta\mu\Delta\rho$ as determined by the Laplace pressure difference across it. The height $\ell_w$
then follows from  the condition that the meniscus must meet each side of the wedge at the correct contact angle. Notice that the amplitude of the divergence
vanishes at the filling phase boundary $\theta=\alpha$ consistent with the requirement that the adsorption becomes microscopic for $T<T_f$.

A particular case of complete wedge filling occurs when the walls are completely wet, $\theta=0$ or completely dry $\theta=\pi$ ( if one studies the wedge-liquid
interface). In this case there are also singular next-to-leading order contributions to the divergence, such that \cite{rascon05}
\begin{equation}
\ell_w\approx\frac{\gamma(\sec\alpha-1)}{\delta\mu\Delta\rho}+\frac{\sec\alpha}{1-\beta_{\rm s}^{\rm co}}\ell_{\pi} +\cdots \label{CoFilling2}
\end{equation}
where $\ell_{\pi}\approx\delta\mu^{-\beta_s^{co}}$ is the thickness of the complete wetting layer at a planar wall-vapour interface (or wall-liquid in the case
of drying). The character of this next-to-leading order correction therefore does depend on the range of the intermolecular forces since these determine the
wetting layer thickness. For systems with short-ranged forces recall that $\ell_\pi\approx -\xi_b\ln\delta\mu$, where $\xi_b$ is the correlation length of the
bulk phase adsorbed at the wall, i.e. $\beta_s^{co}=0$. Strictly speaking this is a mean-field result but, in three dimensions (which is the upper critical
dimension for short ranged fores), interfacial fluctuation effects do not alter this in any significant way, only altering the amplitude by a factor $1+\omega/2$
where $\omega=k_B T/4\pi \gamma \xi_b^2$ is the wetting parameter \cite{dietrich}. For long-ranged intermolecular potentials on the other hand the exponent
$\beta_s^{co}=1/(p+1)$ with $p=2,3$ for non-retarded and retarded dispersion forces respectively \cite{dietrich}. The critical amplitude of the correction term
is similar to that describing the well known Derjaguin correction to the Kelvin equation for capillary condensation in a slit geometry \cite{derj} Only for the
case of short-ranged forces does the correction term have a simple geometrical interpretation arising from the wetting layer along the walls far from the apex.

\begin{figure}[ht]
\includegraphics[width=8.5cm]{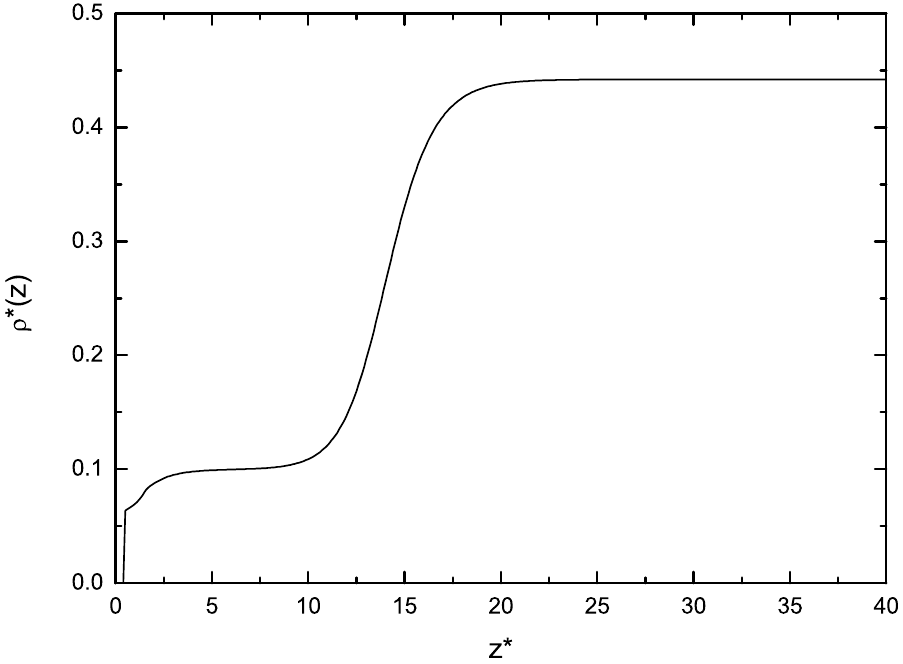}
\caption{Density profile of a liquid near a planar hard wall for $T=0.92\,T_c$ and $\rho_b^*-\rho_l^*=10^{-6}$. Notice the presence of a thick drying layer of
vapour whose density is close to that of the bulk gas. The density is lower at the wall consistent with the pressure sum-rule.}\label{pT13_drho1d-6}
\end{figure}

In this section we test the effective Hamiltonian prediction (\ref{CoFilling2}) for the case of complete drying by vapour at a right angle hard wall wedge. We
suppose the wedge is in contact with a bulk liquid at chemical potential $\mu>\mu_{\rm sat}$. Then, as coexistence is approached from above a bubble of low
density vapour forms at the corner whose height from the apex should be described by Eq.\,(\ref{CoFilling2}). In our calculations we fix the temperature to
$T=0.92\,T_c$ for which bulk coexistence occurs at $\mu_{\rm sat}=-3.96511\epsilon$. As a check of our 2D DFT numerical algorithm, we first studied the planar
hard-wall liquid interface.  In this case, we fixed the  particle density at $z=L$ to a bulk density $\rho_b$, which is slightly higher than the density of the
liquid at saturation, $\rho_l=0.43148\,\sigma^3$. In Fig.\,\ref{pT13_drho1d-6} we display a typical equilibrium density profile $\rho(z)$ showing a fairly thick
drying layer of low density vapour. For comparison the bulk density of gas at this temperature is $\rho_g=0.1\,\sigma^3$. The density near the wall falls and at
contact, $\rho_w\equiv\rho(\sigma)$ should be exactly given by the sum-rule $p=k_BT\rho_w$ where $p$ is the bulk pressure. The measured value of the contact
density with this grid size determines the bulk pressure with an error less than $0.01$\%.

\begin{figure}[h]
\includegraphics[width=8.5cm]{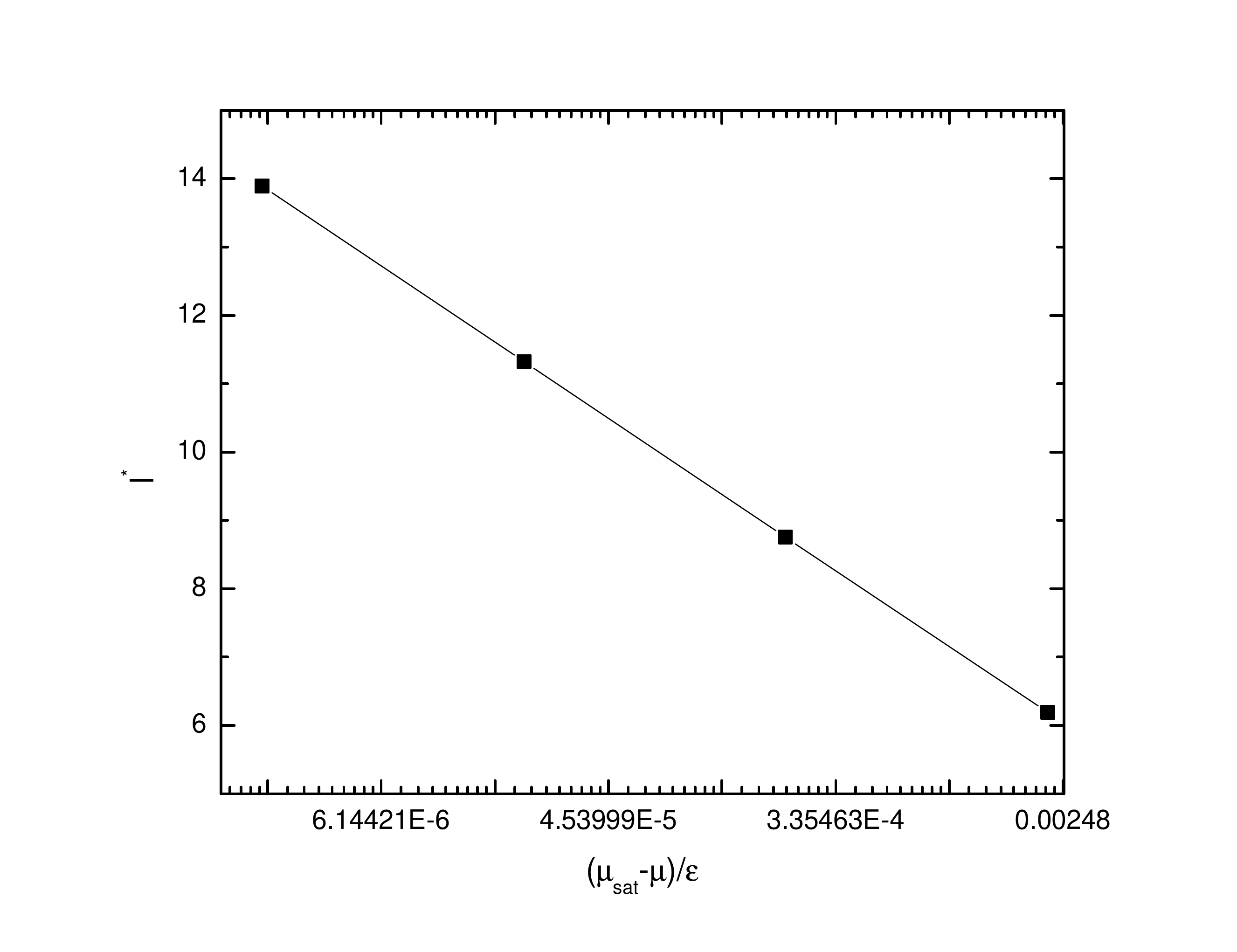}
\caption{Plot of the film thickness $\ell^*_\pi=\ell_\pi/\sigma$ of the adsorbed layer of gas at a hard wall versus the over saturation, measured on a log scale,
for $T=0.92\,T_c$. The slope of the straight line determines the bulk gas correlation length.}\label{l-dmu_T=13_hw}
\end{figure}

From the equilibrium  density profile we determine the adsorption $\Gamma=\int_\sigma^{L}(\rho(z)-\rho_b)\dd z$ and from this the film thickness according to the
standard definition $\ell_\pi=|\Gamma|/\Delta\rho$. In Fig. \ref{l-dmu_T=13_hw} we show the dependence of $\ell_\pi$ on the supersaturation. This  is in
excellent agreement with the expected logarithmic divergence,  for this cut-off LJ fluid, and allows us to identify the bulk correlation length of the gas phase
$\xi_b^{gas}=1.11\,\sigma$. This agrees with the value obtained independently from the decay of the density for the wall-gas interface.
\begin{figure}
\includegraphics[angle=0, width={0.2\textwidth}]{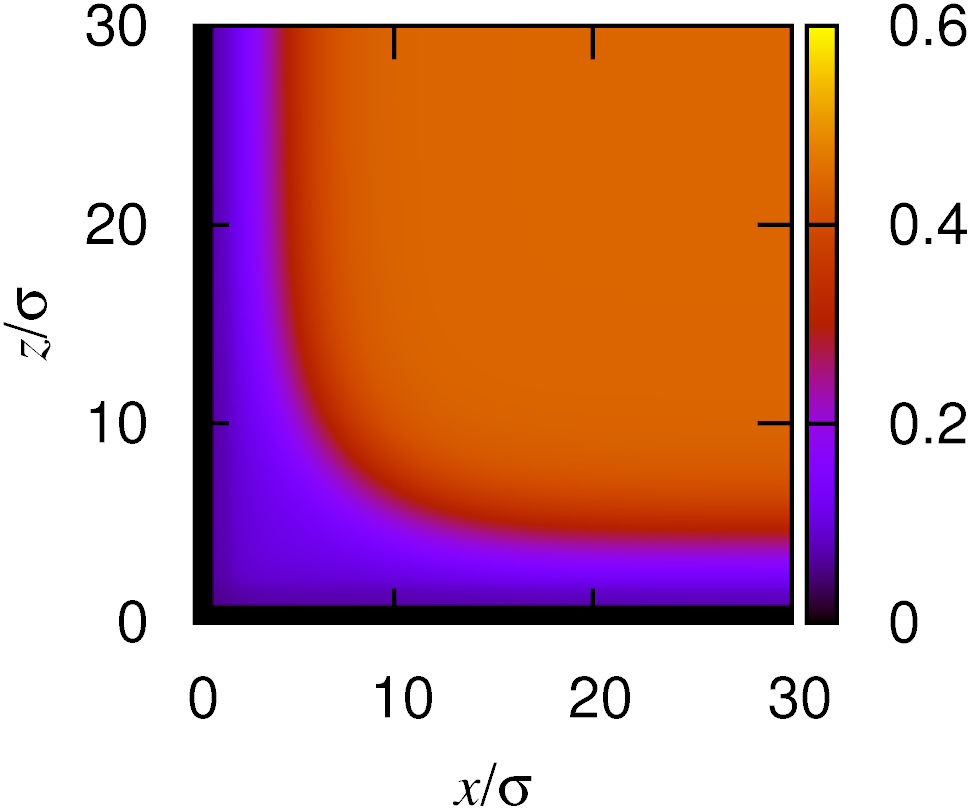}  \includegraphics[angle=0, width={0.2\textwidth}]{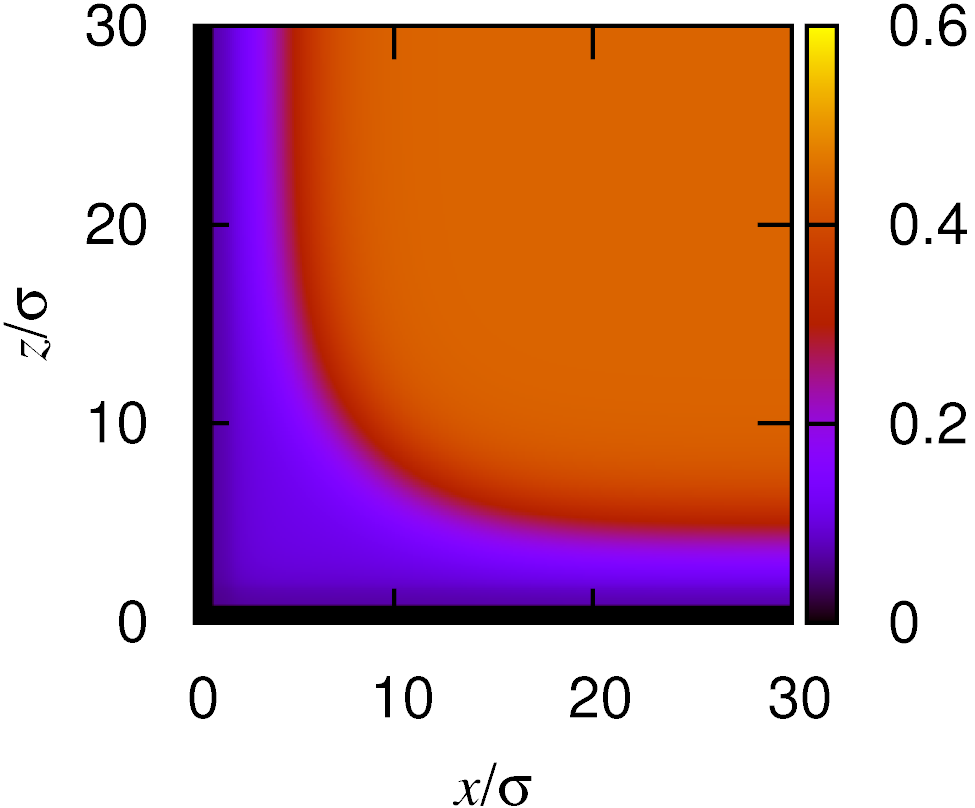}\\
\includegraphics[angle=0, width={0.2\textwidth}]{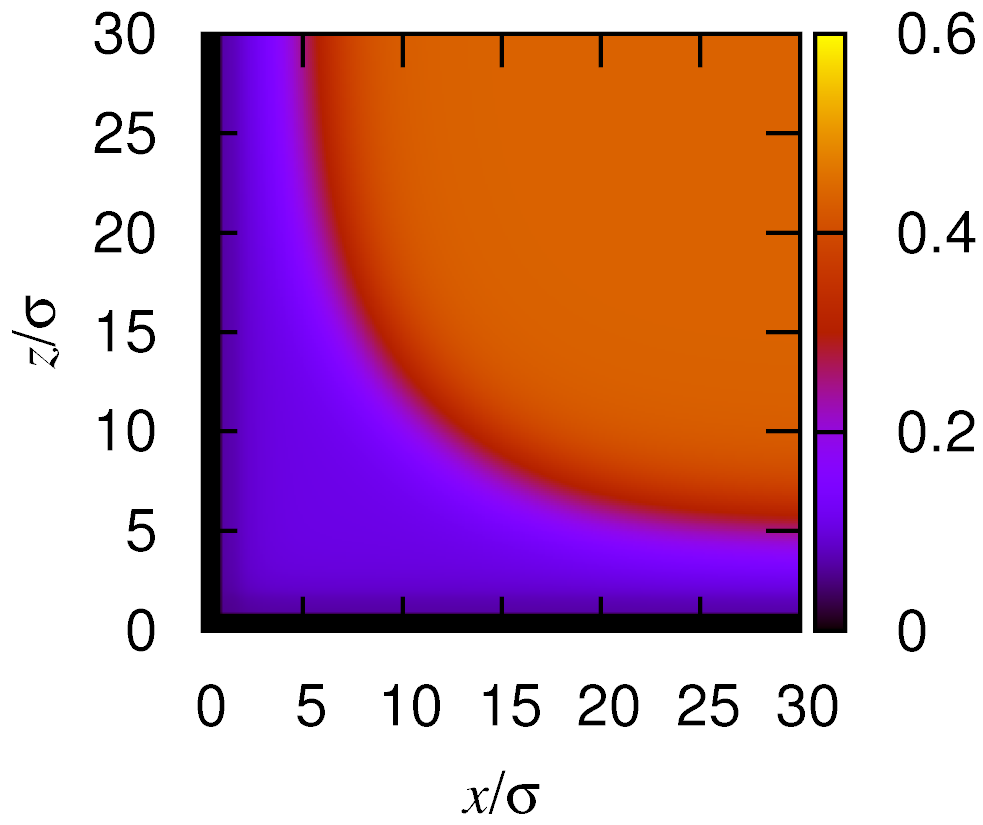} \includegraphics[angle=0, width={0.2\textwidth}]{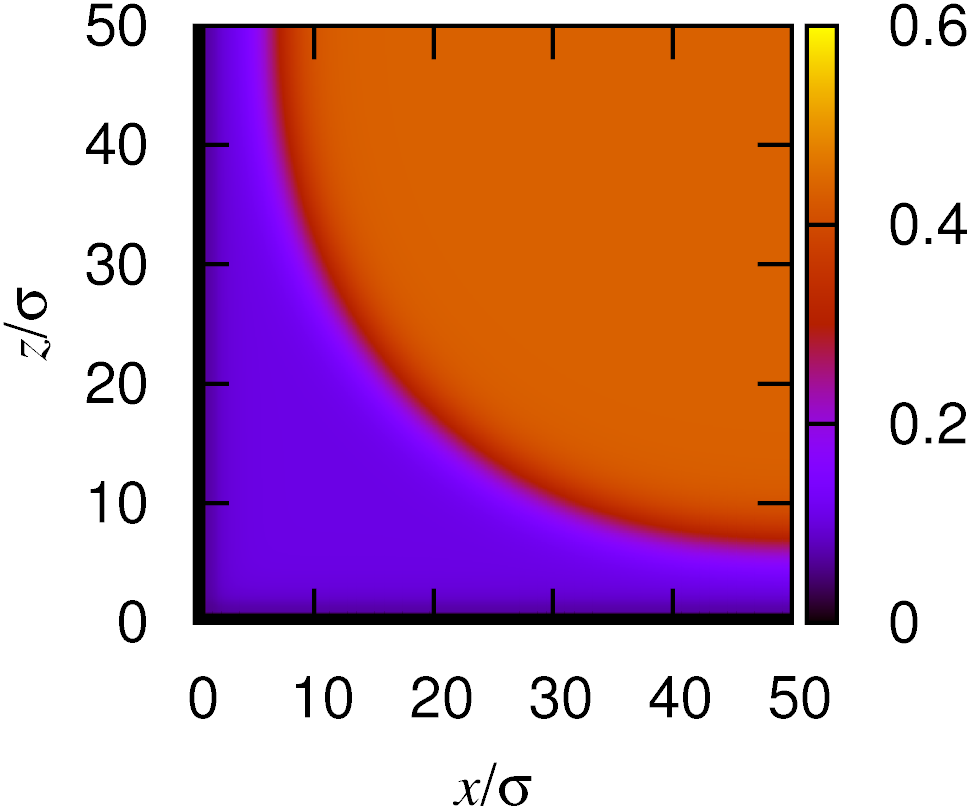}
\caption{Fluid density profiles near a hard wedge in contact with a bulk liquid at temperature $T=0.92\,T_c$. The bulk density differences
($\delta\rho\equiv\rho_b-\rho_l$) from top left to right bottom are: $\delta\rho\sigma^3=0.01, 0.008, 0.005$, and $0.003$, respectively.}\label{hwed_profs}
\end{figure}

We now turn our attention to the hard wall wedge geometry, corresponding to the external potential (\ref{hard_wedge}), and numerically study the interface with a
bulk liquid for different chemical potentials approaching bulk coexistence. In Fig.\,\ref{hwed_profs} we show four different 2D  density profiles $\rho(x,z)$ for
values of the chemical potential progressively closer to saturation. For the values of the chemical potential chosen one can see qualitatively that far from the
wedge apex the adsorption of gas is rather small corresponding to thin drying films only a few $\sigma$ thick. These are  indicative of the planar wall-liquid
interface. In contrast, even for the largest value of the chemical potential the geometry enhanced preferential adsorption of vapour at the apex, via the
formation of a meniscus, is clearly apparent. Upon approaching saturation, the distance $\ell_w$ of the meniscus from the apex,  increases, and diverges as
$\mu\to\mu_{\rm sat}$. This divergence is far stronger than the logarithmic increase of the drying film at a planar wall.

\begin{figure}[h!]
 \includegraphics[angle=0, width=7.5cm]{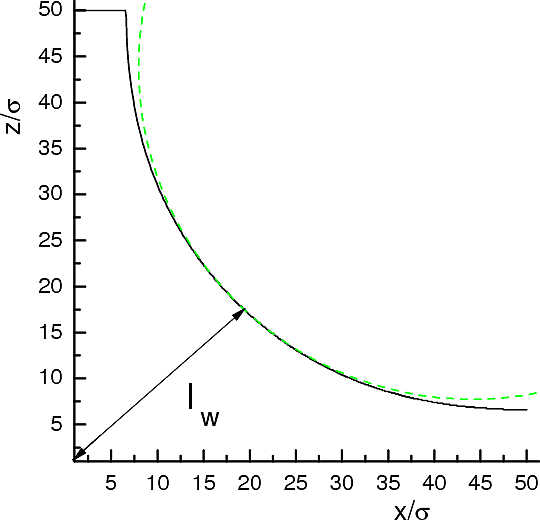}
\caption{Contour of the meniscus between the vapour and liquid phases in a hard wedge at temperature $T=0.92\,T_c$ and bulk density
$\rho_b-\rho_l=0.003\,\sigma^3$. Also shown for comparison is a circular meniscus of Laplace radius $R=\frac{\gamma}{\delta\mu\Delta\rho}$.}\label{i_3d-3}
\end{figure}

\begin{figure}[h]
    \includegraphics[angle=0, width=8.5cm]{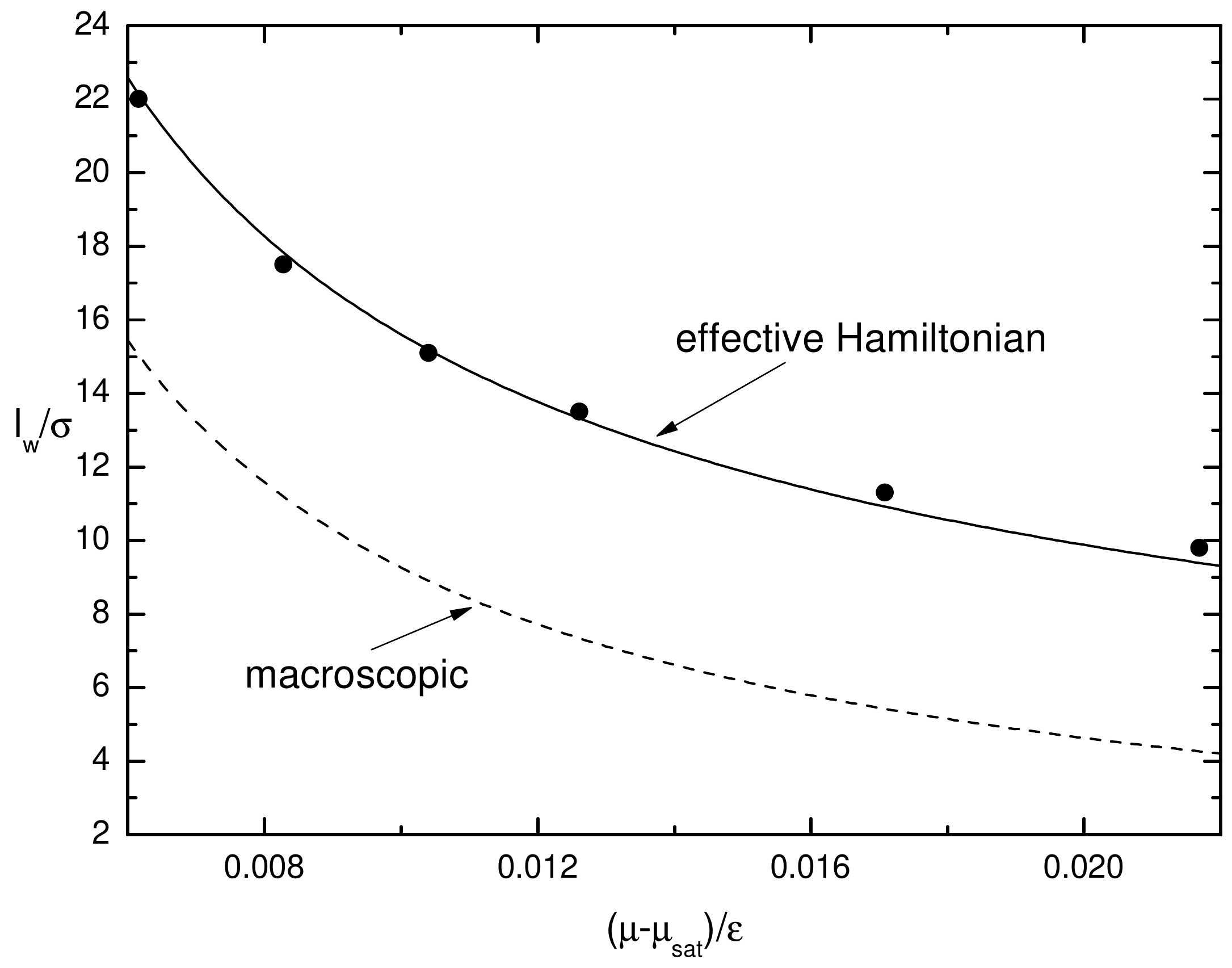}
\caption{Numerical results (symbols) for the divergence of the meniscus filling height $\ell_w$, shown in comparison with the macroscopic expression
$\ell_w=\frac{\gamma(\sec\alpha-1)}{\delta\mu\Delta\rho}$ (dashed curve) and the effective Hamiltonian prediction (\ref{CoFilling2}) (solid curve) which includes
the logarithmic next-to-leading order correction.}\label{l_delta_mu_log}
\end{figure}

In Fig.\,\ref{i_3d-3} we show a numerically determined meniscus shape corresponding to the loci of the local mid-point interfacial density where
$\rho(x,z)=\frac{\rho_b+\rho_l}{2}$ for the case  where  $\rho_b=\rho_l+0.003\,\sigma^3$. The shape of the meniscus is very nearly circular,  as can be seen from
comparison with the green circle which has the Laplace radius $R=\gamma/\delta\mu\Delta\rho=36.3\,\sigma$ for this particular chemical potential. This gives us
some confidence that even for the present small system sizes the predictions of macroscopic and effective Hamiltonian theory are still valid. Finally, and most
importantly, the numerically determined divergence of the  filling height, $\ell_w$, is shown in Fig.\,\ref{l_delta_mu_log} (symbols). The  dashed  curve is the
macroscopic theoretical expression  which corresponds to just the first term in  (\ref{CoFilling2}). For $T=0.92\,T_c$, this corresponds to the curve $
\ell_w=0.09263\,\sigma\varepsilon/\delta\mu$. The shape of this is very similar to the numerical results for the film thickness but lies systematically below it.
The solid curve is the theoretical result now allowing for the next-to-leading order correction in (\ref{CoFilling2}), which recall is logarithmic for the
present cut-off LJ fluid. We emphasize, since the agreement is so good, that this is not a fit and there are no adjustable parameters. Effective Hamiltonian
theory therefore gives an excellent quantitative description of complete filling at a hard wall wedge. One simplifying feature of the hard wall in contact with a
vapour wetting or drying layer, of course, is that there are no packing effects to worry about. Filling by liquid is potentially more complicated because of such
effects. We turn to this in the next section in the context of the filling transition itself.

\subsection{Filling with long-range wall-fluid forces}

\begin{figure}
\includegraphics[width=0.45\textwidth]{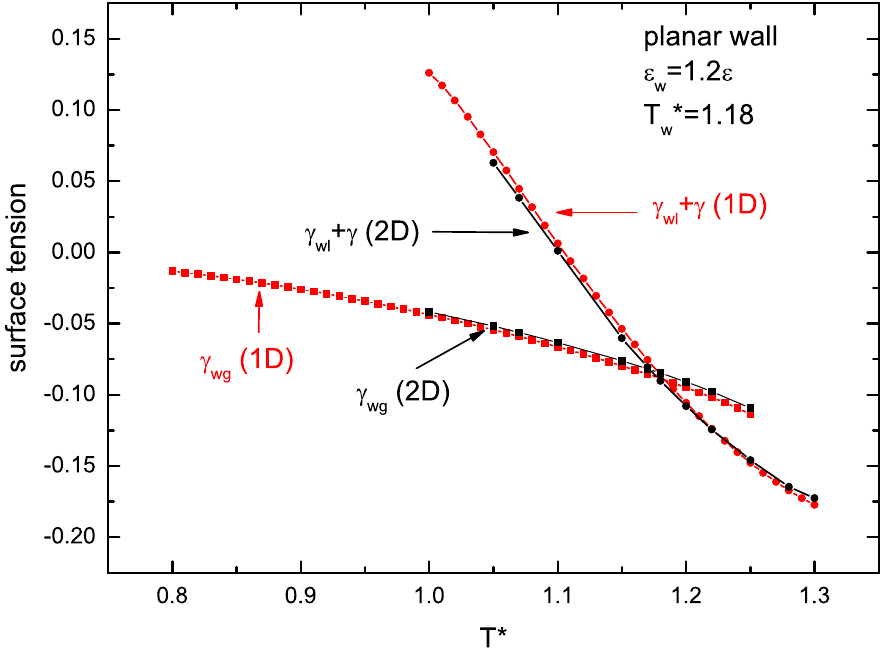}\\
\includegraphics[width=0.45\textwidth]{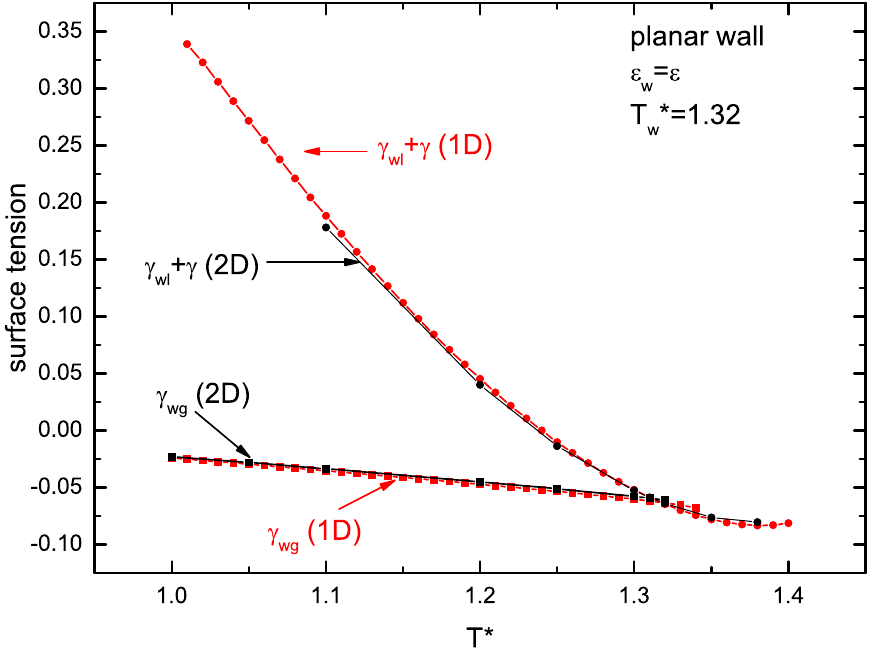}\\
\includegraphics[width=0.45\textwidth]{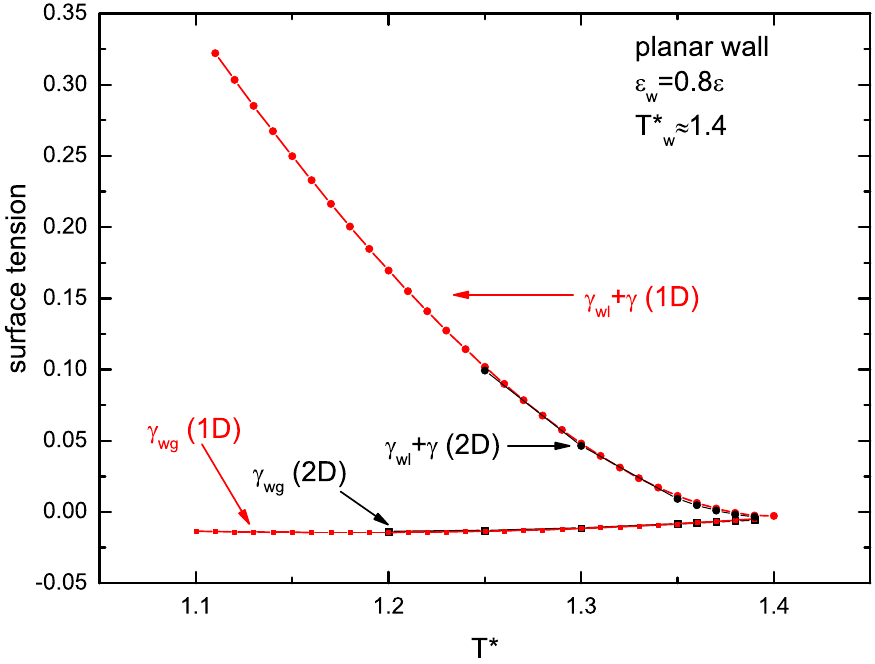}
\caption{Determination of the wetting temperature $T_w$ from the intersection of the wall-gas tension $\gamma_{wg}$ and summed tensions $\gamma_{wl}+\gamma$. Results obtained from 1D calculations (in red) are compared with
those of the 2D calculation (in black).}\label{wet}
\end{figure}

\begin{figure}
 \includegraphics[width=0.45\textwidth]{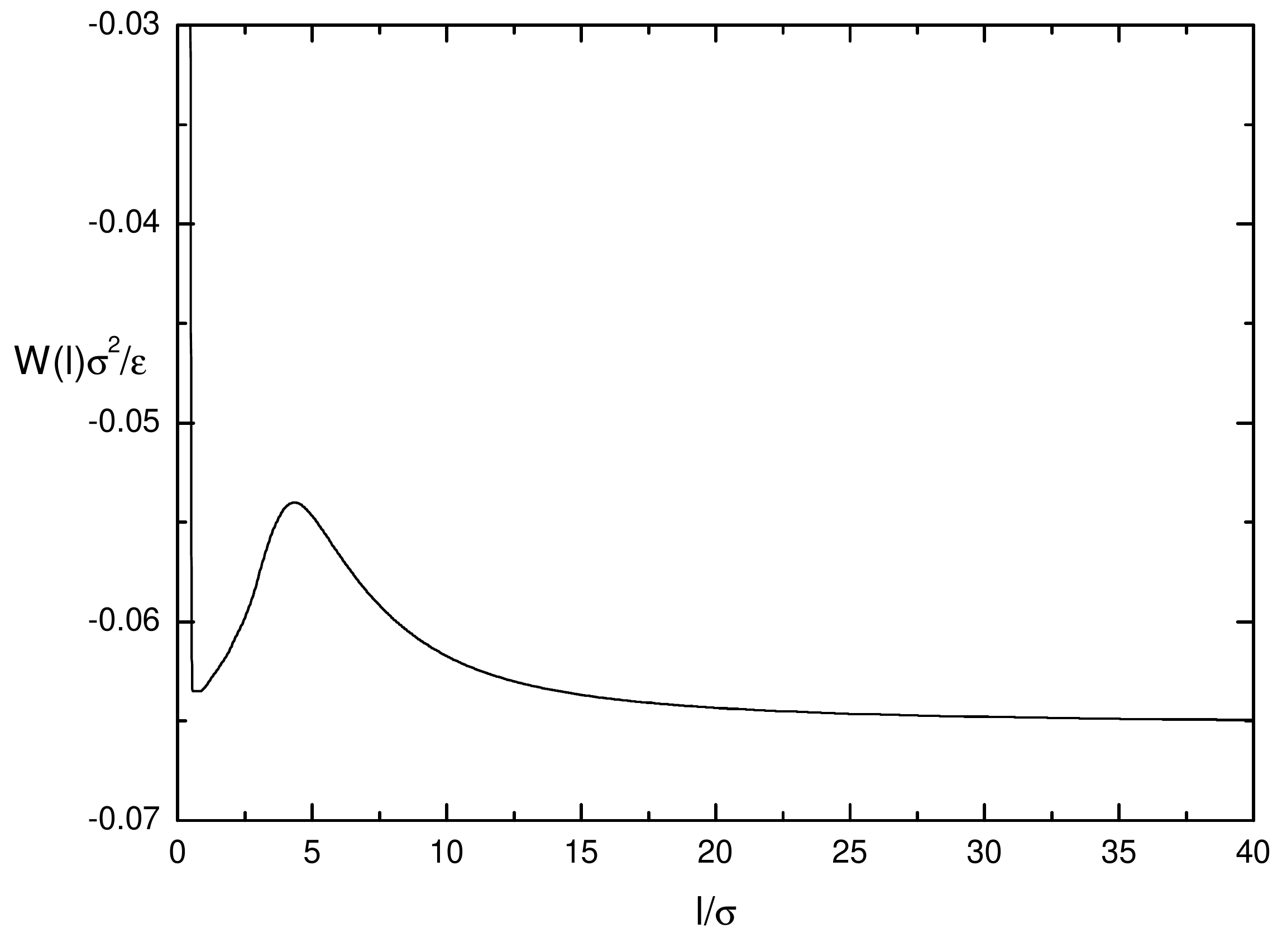}
 \includegraphics[width=0.45\textwidth]{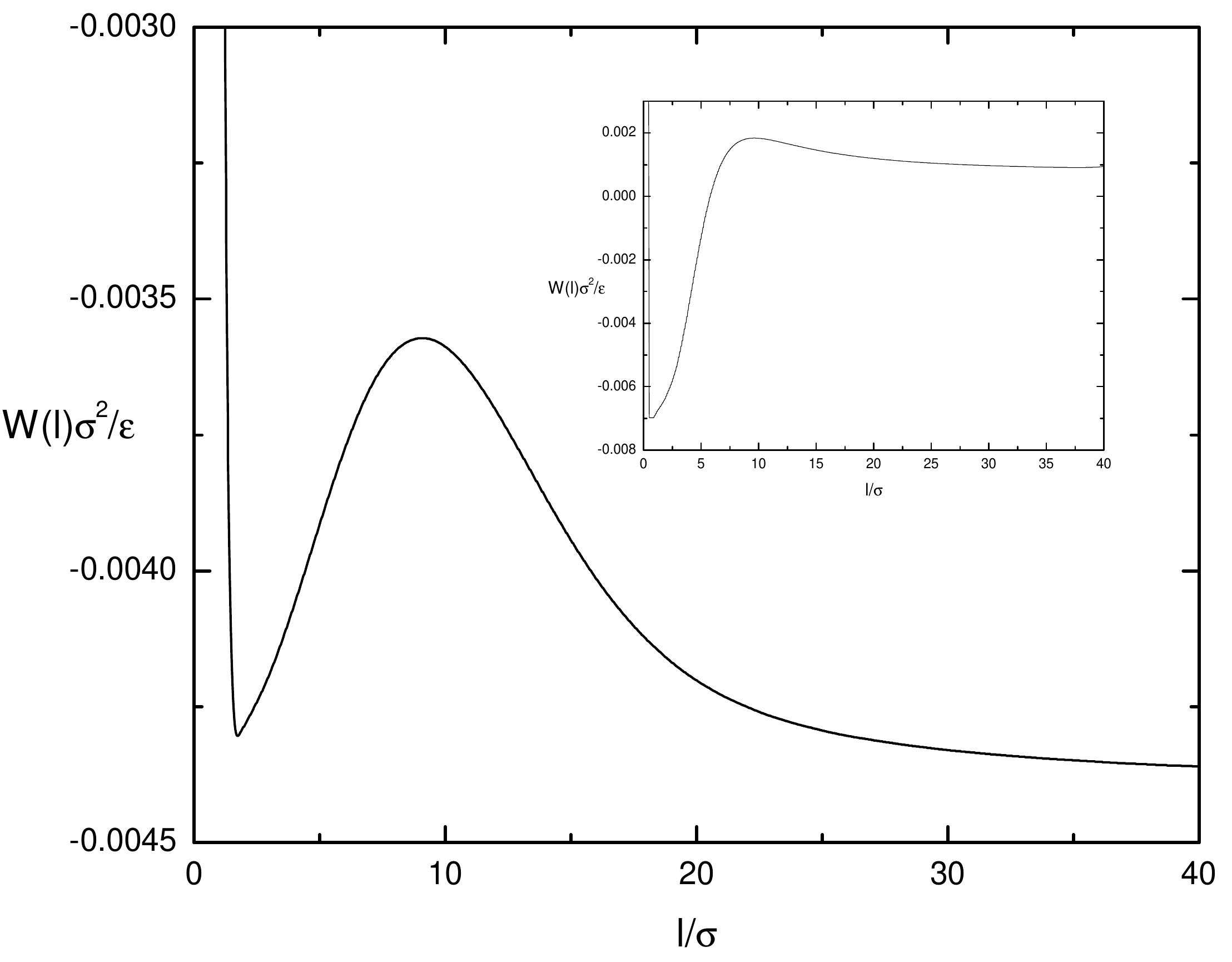}
\caption{Interfacial binding potentials $W(\ell)$, in dimensionless units, as a function of the adsorbed film thickness. The upper panel corresponds to $\varepsilon_w=\varepsilon$ and for a temperature close to $T_w^*=1.32$.
Notice the presence of a prominent activation barrier at $\ell_B\approx 4\,\sigma$. In tho lower panel we show the binding potential for $\varepsilon_w=0.8\,\varepsilon$ also close to its wetting transition at $T_w^*=1.4$. In
this case, the activation barrier is an order of magnitude smaller and situated further from the wall at $\ell_B\approx 10\,\sigma$. In the inset we show the binding potential at a lower temperature which is close to the
filling temperature $T_f^*\approx 1.38$ for this same system. Notice that an activation barrier is still present. In all cases the results correspond to a bulk coexistence.}\label{bind08}
\end{figure}


In this section we go beyond the pure hard wall wedge and turn on the long-range wall-fluid attraction. We consider three different interaction strengths; (i) $\varepsilon_w=1.2\,\varepsilon$, (ii) $\varepsilon_w=\varepsilon$
and (iii) $\varepsilon_w=0.8\,\varepsilon$. For each of these we first consider the corresponding planar wall-fluid interfaces and determine the temperature dependence of the contact angle $\theta(T)$  from the wall-gas and
wall-liquid surface tensions using Young's equation $\cos\theta=(\gamma_{wg}-\gamma_{wl})/\gamma$. Each of these systems exhibits a wetting transition by liquid at a wall-vapour interface. As expected, these transitions are
first-order since the wall-fluid and fluid-fluid forces have different ranges. In Fig.\,\ref{wet} we show the numerically determined value of the wetting temperatures using both the 1D and 2D DFT calculations. The crossing of
the wall-gas tension $\gamma_{wg}$ and summed tensions $\gamma_{wl}+\gamma$ gives consistent values of $T_w^*=1.18$ (or $T_w=0.83\,T_c$), $T_w^*=1.31$ (or $T_w=0.93\,T_c$) and $T_w^*=1.4$ (or $T_w=0.99\,T_c$) for the cases
(i)--(iii), respectively. For the strongest wall-fluid interaction, $\varepsilon_w=1.2\,\varepsilon$ the wetting transition is strongly first-order as can seen from the crossing of the free-energy branches. To further
emphasise this  we have determined numerically the interfacial binding potential $W(\ell)$ representing the excess free energy of a wetting film constrained to be of thickness $\ell$. For $\varepsilon_w=\varepsilon$ this is
shown at the upper panel of Fig.\,\ref{bind08} for a temperature close to $T_w$. As can be seen there is a clear activation barrier located near $\ell_B\approx 4\sigma$.

For the weakest wall-fluid potential $\varepsilon_w=0.8\,\varepsilon$, where $T_w$ is very close to the bulk critical temperature the transition is weakly first-order as can be seen from the near tangential meeting of the
surface tensions. This is more apparent when one numerically determines the interfacial binding potential for $T\approx T_w$ for this interaction strength (see lower panel of Fig.\,\ref{bind08}) by minimizing the grand
potential subject to a constraint of fixed film thickness \cite{hend_bind}. This function still exhibits an activation barrier but this is an order magnitude smaller than for the $\varepsilon_w=\varepsilon$ case and its
location near $\ell_B\approx 10\,\sigma$ is far further from the wall. In set we show the binding potential at a lower temperature (at the filling transition) which we will return to this later.

\begin{figure}
\includegraphics[width=0.5\textwidth]{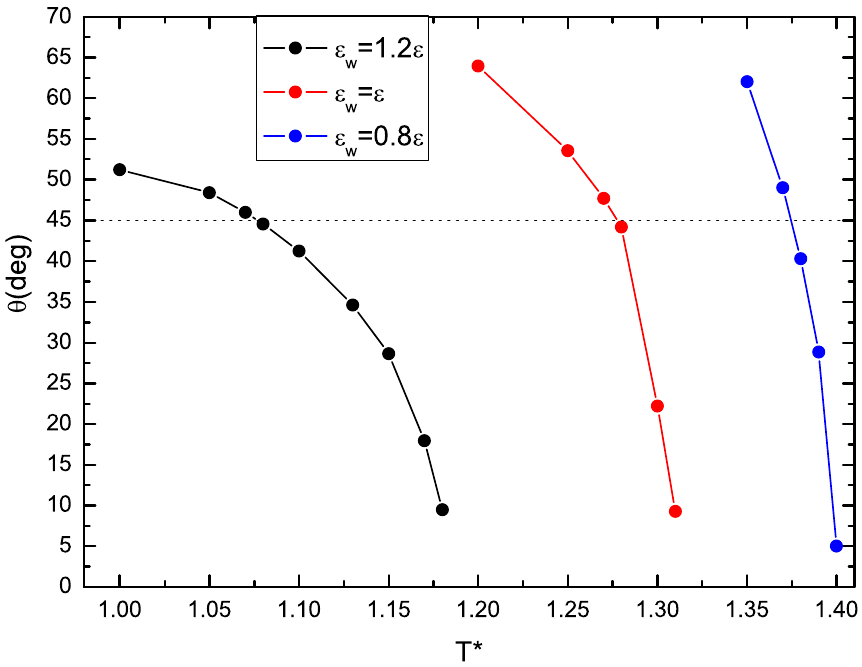}
\caption{Variation of the contact angle with temperature for three different wall strengths. The intersection with the dashed line at $\theta=45^\circ$ is the
thermodynamic prediction for the filling temperature for each system. }\label{cont_angle}
\end{figure}

A plot of the contact angles as a function of temperature for each of the wall strengths is shown in Fig.\,\ref{cont_angle} where the intersection with
$\alpha=\pi/4$ gives, from  the thermodynamic prediction (\ref{fill}), the theoretical value of the filling transition in a right angle wedge. These are
$T_f^*=1.075$ (or $T_f=0.76\,T_c$), $T_f^*=1.275$ (or $T_w=0.90\,T_c$) and $T_f^*=1.375$ (or $T_w=0.97\,T_c$) for interaction strengths (i)--(iii) respectively.


\begin{figure}
\includegraphics[width=0.5\textwidth]{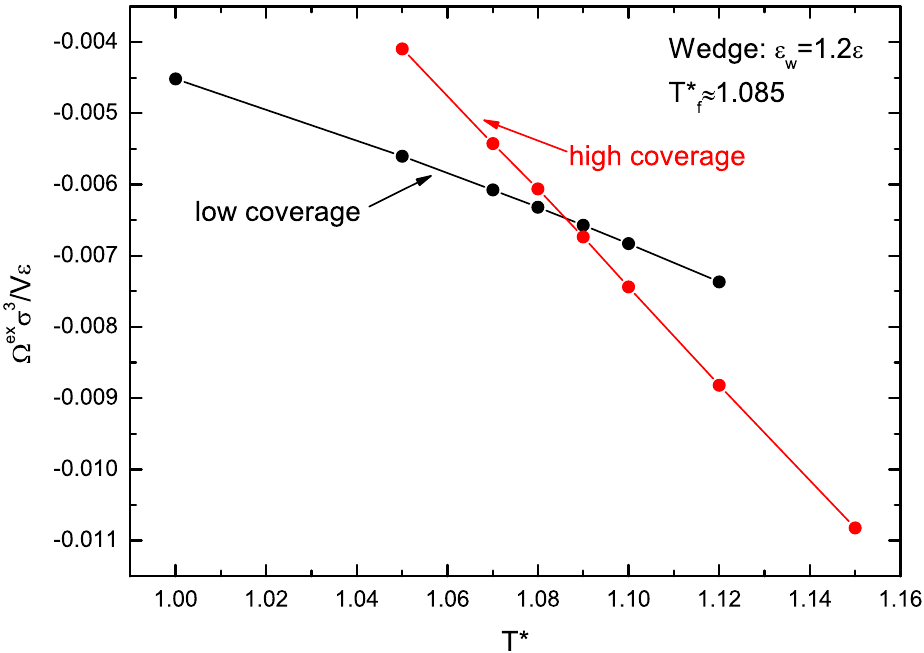}
\caption{Location of a first-order filling transition from the crossing of two separate free energy branches corresponding to microscopic and macroscopic
adsorptions in the wedge geometry for $\varepsilon_w/\varepsilon=1.2$. Here $V$ is the available volume which is the length of the wedge multiplied by
$(L-\sigma)^2$.}\label{w_12}
\end{figure}

\begin{figure}
\includegraphics[width=0.5\textwidth]{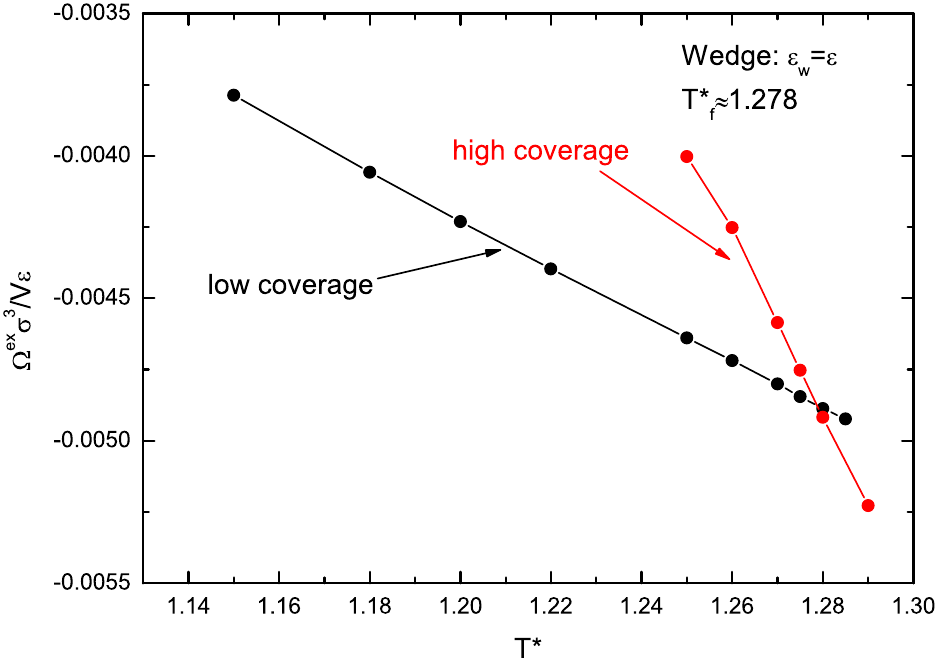}
\caption{Location of a first-order filling transition from the crossing of two separate free energy branches corresponding to microscopic and macroscopic
adsorptions in the wedge geometry for  $\varepsilon_w=\varepsilon$. }\label{w_1}
\end{figure}

\begin{figure}
\includegraphics[width=0.4\textwidth]{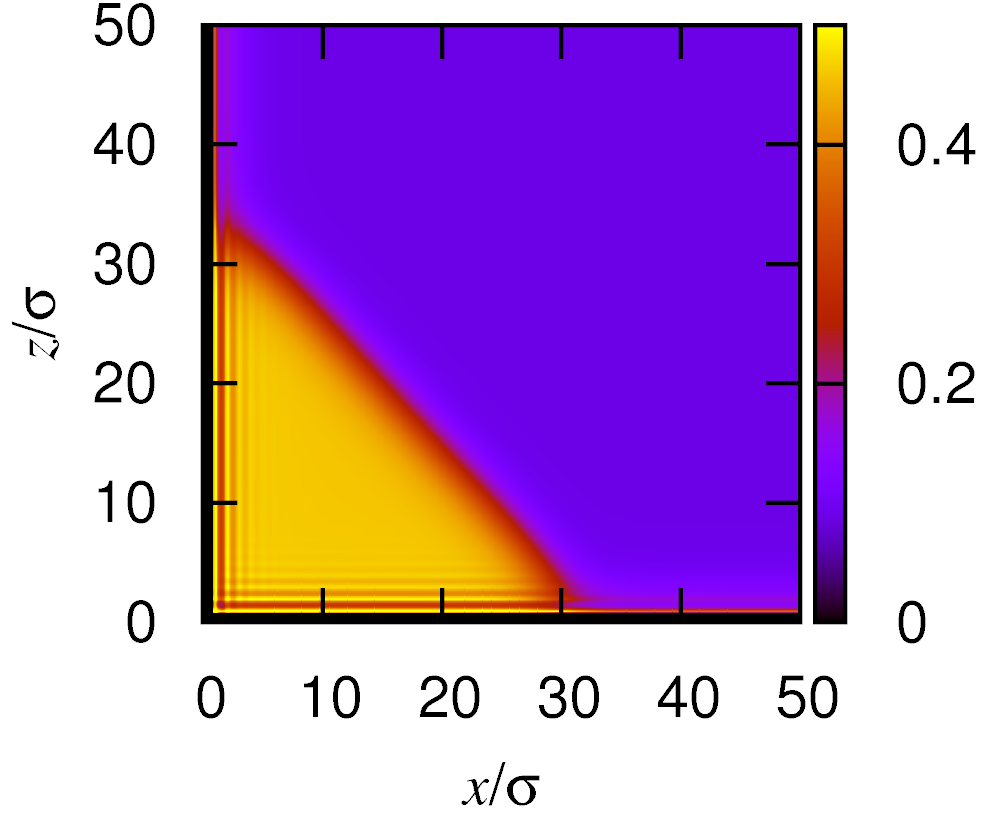}
\includegraphics[width=0.4\textwidth]{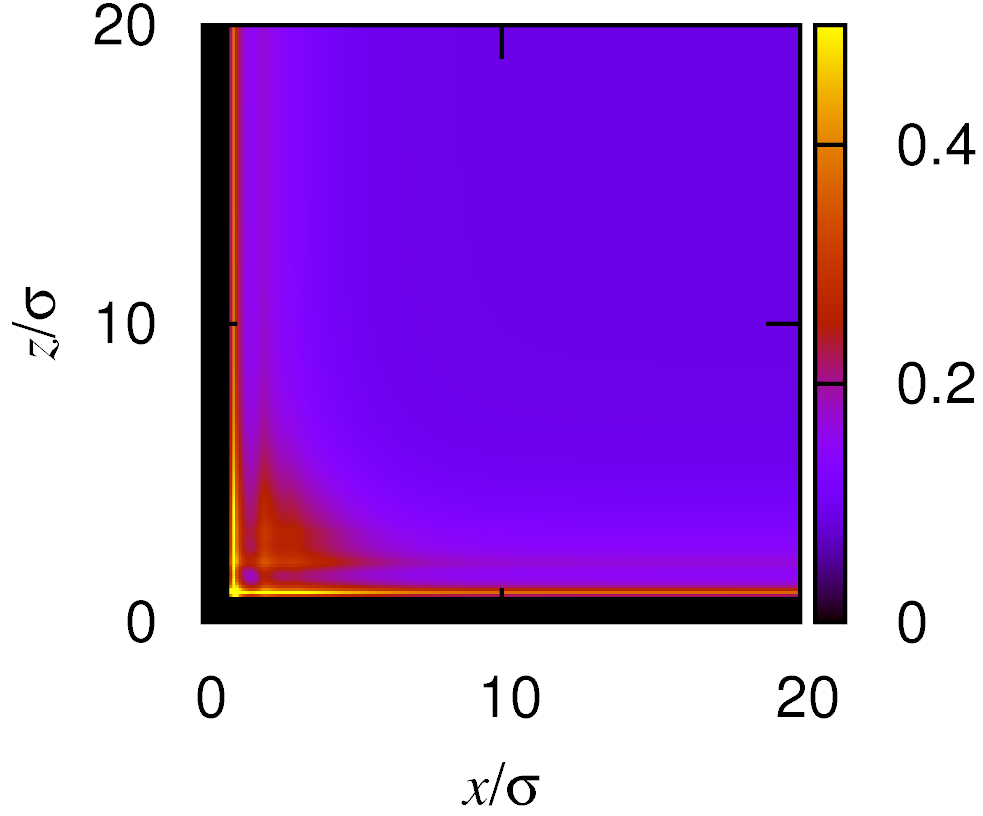}
\caption{Coexisting density profiles for wedge-vapour interfaces at a first-order filling transition for wall strength  $\varepsilon_w=\varepsilon$
(corresponding to $T^*_f=1.28$). The upper panel shows the macroscopic  configuration in which the meniscus is far from the wall and meets each wall at the
contact angle. The lower panel shows the coexisting microscopic configuration in which the interface is tightly bound to the apex.}\label{pw1_1.28}
\end{figure}

\begin{figure}
\includegraphics[width=0.4\textwidth]{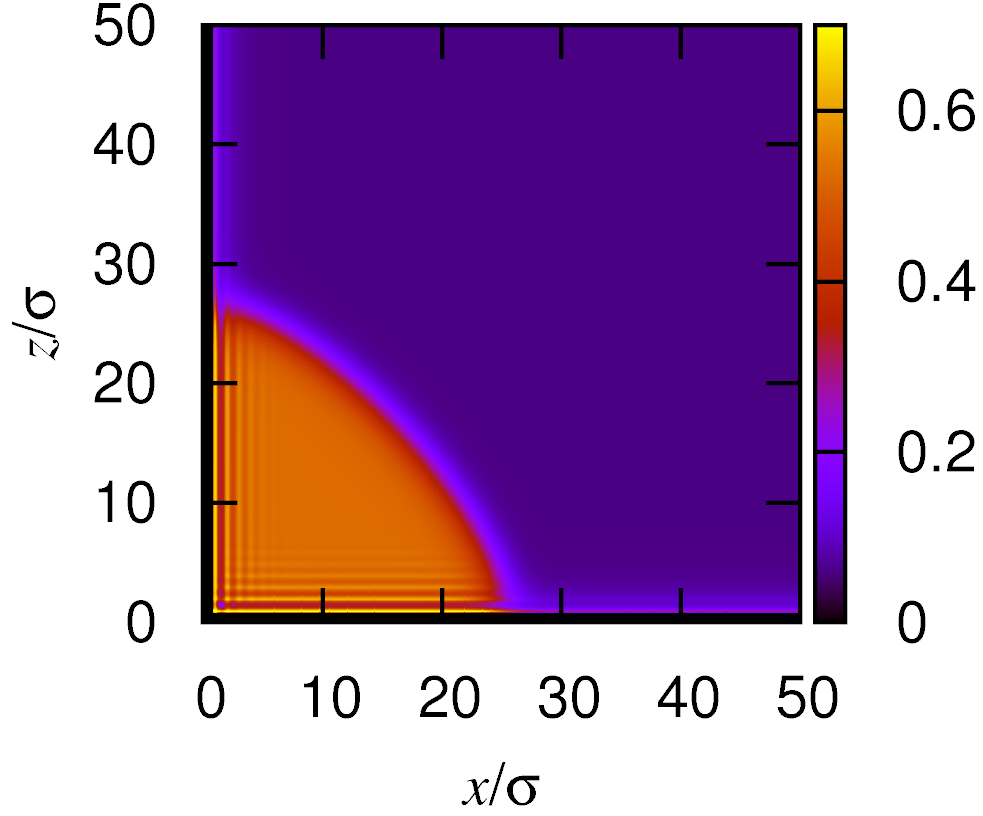}
\caption{Example of a metastable configuration corresponding to a macroscopic meniscus with negative curvature at a temperature $T^*=1.2$ which is below the
filling temperature ($T^*_f=1.28$).}\label{pw1_1.2}
\end{figure}


We now turn attention to the numerical analysis of the equilibrium density profiles and phase behaviour in the wedge geometry. Our first task is to numerically determine the location of any filling transition and compare with
the above theoretical predictions for the three different wall-fluid interaction strengths. To this end we sit at bulk coexistence $\mu=\mu_{\rm sat}$ and  minimize the grand potential $\Omega[\rho]$ to a global or local
minimum $\Omega$, starting from two different initial configurations: a high density liquid and a low density vapour. If the system exhibits a first-order filling transition then we can expect that in the vicinity of $T_f$
these initial configurations will converge to different equilibrium profiles, corresponding to microscopic and macroscopic adsorptions of liquid, respectively. These will  have identical grand potentials at the filling
transition. Obviously finite-size constraints limit the size of the macroscopic liquid layer, the size of which scales with the box area $L^2$.  If, on the other hand, the filling transition is continuous there will be a
unique equilibrium phase. In this case a plot of the total adsorption versus $T$ will have no hysteresis loop but should still show a dramatic continuous increase near $T_f$. Plots of the excess grand potential  $\Omega^{\rm
ex}=\Omega+pV$ per unit length of the wedge obtained in this manner are shown in Figs.\,\ref{w_12} and \ref{w_1}. For the two strongest interaction strengths there are two separate branches of the free-energy, indicating a
first-order filling transition. The crossing of the free-energies yields filling temperatures $T_f^*=1.085$ and $T_f^*=1.278$ for $\varepsilon_w=1.2\,\varepsilon$ and $\varepsilon_w=\varepsilon$, respectively which are close
to the theoretical predictions obtained from $\theta(T_f)=\pi/4$. The slight discrepancy between the values is a consequence of the finite-size limitations of our numerical analysis. In Fig.\,\ref{pw1_1.28} we show the
coexisting density profiles, corresponding to microscopic (lower panel) and macroscopic (upper panel) states, at the filling temperature for wall interaction strength $\varepsilon_w=\varepsilon$.  Notice that the macroscopic
meniscus is nearly flat, as it should be since we are at bulk coexistence and the interface must meet the walls at a contact angle equal to $\pi/4$. Notice that for the microscopic configuration the thickness $\ell_w$ of the
adsorbed layer is larger than the wetting layer thickness (far from the apex) but of the same order as the distance of the activation barrier $\ell_B$ for the corresponding binding potential for the wetting transition, see
inset of Fig.\,\ref{bind08}.  Both microscopic and macroscopic profiles show layering behaviour close to the apex. Also shown in Fig.\,\ref{pw1_1.2} is a metastable configuration for $T<T_f$ representing a macroscopic
adsorption of liquid with a concave meniscus. This curvature is necessary in order that the meniscus meets each wall at the correct contact angle.

\begin{figure}
\includegraphics[width=0.5\textwidth]{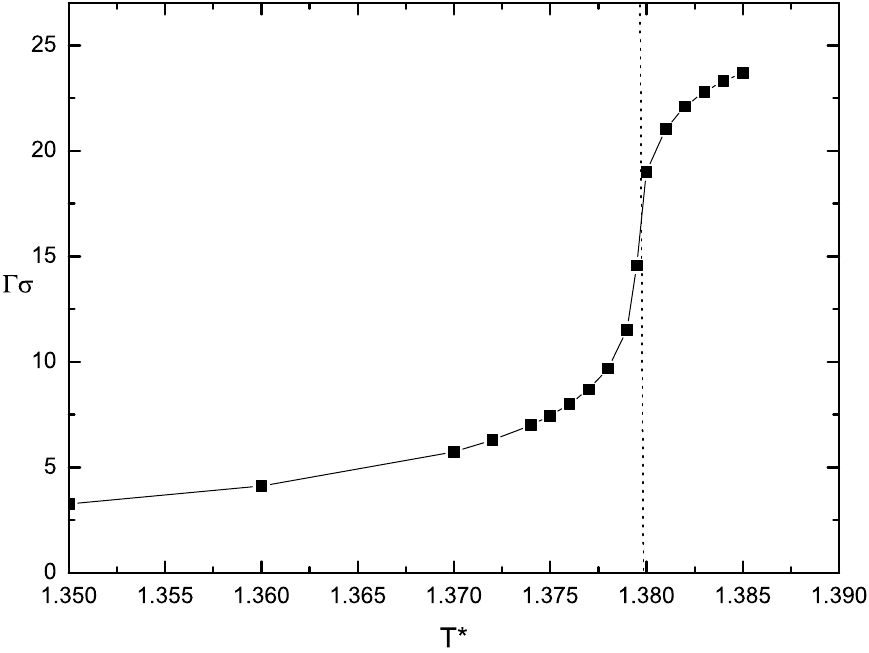}
\caption{Plot of the adsorption (in reduces units) as a function of temperature for the wedge model with the weakest wall interaction
$\varepsilon_w=0.8\,\varepsilon$. The adsorption increases sharply but continuously in the vicinity of the filling temperature indicating that the transition is
continuous. }\label{w_08}
\end{figure}

\begin{figure}
\includegraphics[width=0.5\textwidth]{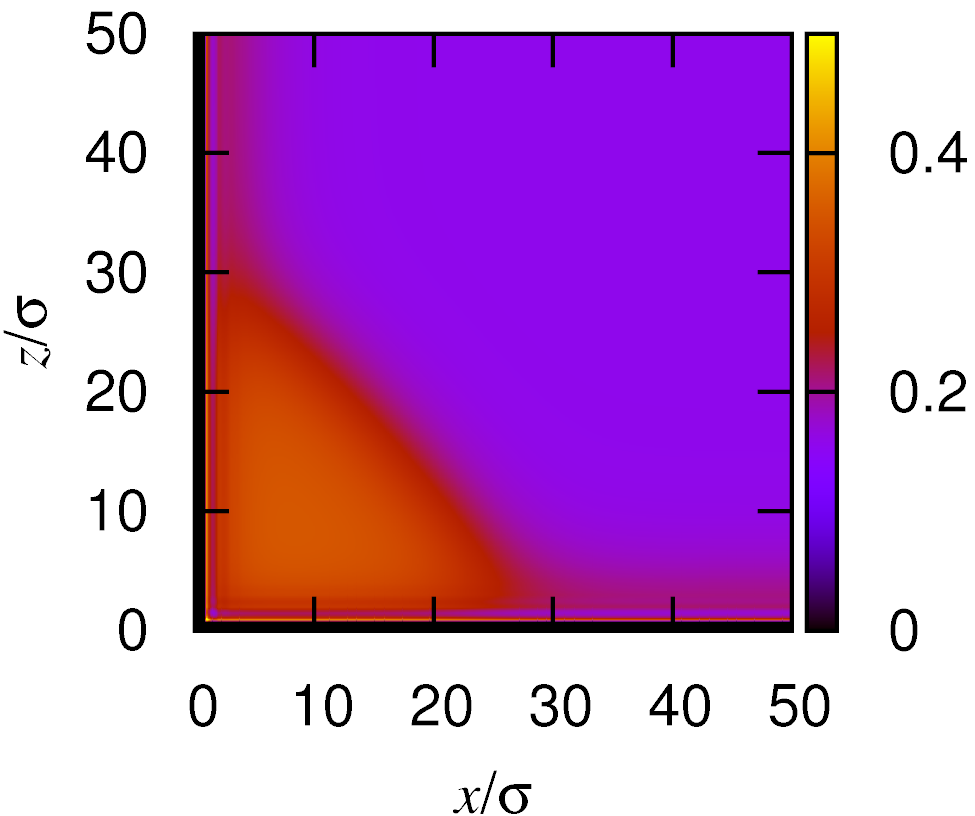}
\caption{Density profile for the weakest wall interaction,
$\varepsilon_w=0.8\,\varepsilon$, near the filling temperature. }\label{pw08_T138}
\end{figure}

Most interestingly, for the weakest wall strength $\varepsilon_w=0.8\,\varepsilon$ we have found that there is only a single branch to the equilibrium grand potential i.e. both high and low density initial coverages converge
to a unique equilibrium phase. This means that either the filling transition is continuous (critical), or so weakly first-order that the present $L\times L$ finite-size grid, is not large enough to see the jump in the
adsorption. A plot of the adsorption $\Gamma=\int\int \dd x\dd z (\rho(x,z)-\rho_b)$ versus $T$ is shown in Fig.\,\ref{w_08}. As is evident, there is indeed a dramatic but continuous increase in the adsorption near the
anticipated  $T_f^*\approx1.38$, indicating that a continuous filling or possibly finite-size rounded weakly first-order filling transition is taking place. Strong evidence that this a genuine critical filling transition comes
from two sources. Firstly, consider the unique density profile at $T=T_f$ shown in Fig.\,\ref{pw08_T138}. The thickness of the meniscus $\ell_w$ is much larger than the length-scale $\ell_B\approx 10\,\sigma$ associated with
the wetting activation barrier. If finite size rounding was an issue we would expect that $\ell_w<\ell_B$ or at least that these length-scales would be comparable. We also emphasise here that even though we are quite close to
the bulk-critical temperature (recall, $T_f\approx 0.975\,T_c$), the bulk (liquid) correlation length is still of the order of $\sigma$ and is much smaller than the overall meniscus size. This is also clear in
Fig.\,\ref{pw08_T138} where the width of the interface, separating liquid from gas, is much smaller than $\ell_w$. This clearly indicates that the mean-field character of our DFT should not play any significant role regarding
the location and the order of the transition. Secondly, we can compare with effective Hamiltonian theory for the meniscus thickness $\ell_w$ and adsorption at critical filling. This predicts that, in an infinite wedge, these
diverge as \cite{wood2}
 \bb
\ell_w\sim(T_f-T)^{-\beta_w}\,,\hspace{1cm} \Gamma\sim(T_f-T)^{-2\beta_w\,,} \label{crit_fill}
 \ee
where the adsorption is simply the square of the film thickness owing to the triangular shape of the meniscus. The mean-field value of the critical exponent $\beta_w=1/p$ and, incidentally, is not altered by interfacial
fluctuation effects in three dimensions provided $p<4$, see Ref. \cite{wood2}. In our model $p=2$ so we should expect that if the filling transition is continuous the adsorption increases as $\Gamma\sim(T_f-T)^{-1}$ on
approaching the filling temperature. In our final figure we show a log-log plot for the growth of the adsorption for $T<T_f$, in which we use the numerical estimate of the filling temperature $T^*_f=1.38$. From this we
estimate $\beta_w=0.46\pm0.05$ which is in a reasonably good agreement with the effective Hamiltonian prediction.


\begin{figure}
\includegraphics[width=0.5\textwidth]{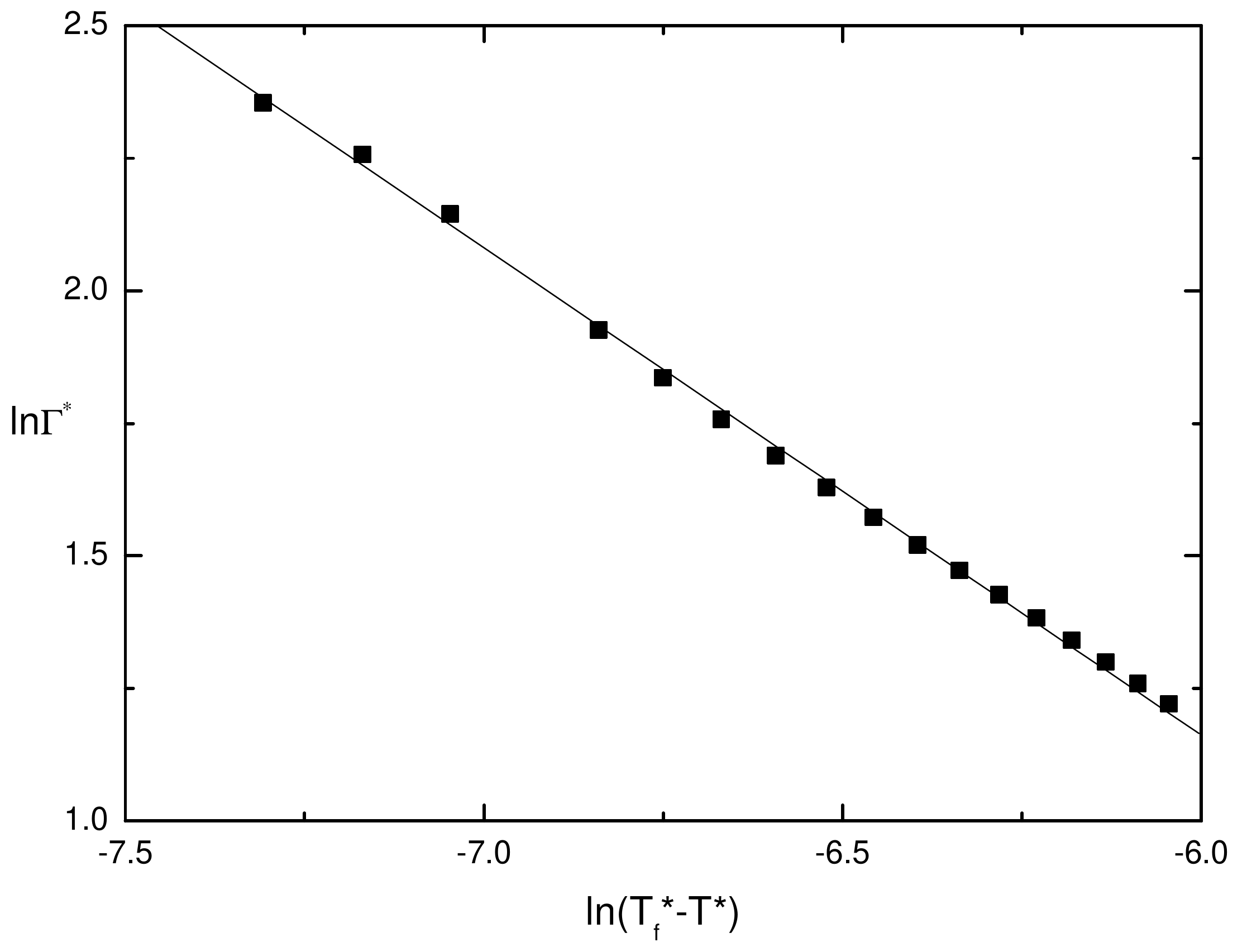}
\caption{Log-log plot of the adsorption vs the scaling field $T_f-T$ for the weakest wall interaction strength $\varepsilon_w=0.8\,\varepsilon$. The slope of the
straight line is $-0.92$. }\label{w_08_log}
\end{figure}

\section{Discussion}

In this paper we have presented our results of numerical studies of complete, first-order and critical filling transitions in a rectangular wedge using a
non-local density functional theory. To the best our knowledge this is the first time that filling transitions have been studied using modern microscopic DFT and
our work complements earlier effective Hamiltonian, square gradient and simulation studies. For the case of complete filling the results of the DFT confirm
effective predictions for leading and next-to-leading order critical exponents and amplitudes to a remarkable accuracy. However we believe our most important
finding is that close to the bulk critical temperature the wedge filling transition is continuous even though the walls themselves exhibit first-order wetting.
Crucially this occurs in the presence of realistic long-ranged wall-fluid interactions (and for a cut-off LJ fluid), which is the system that is most accessible
experimentally. To end our paper we discuss this in more depth.


The change from first-order to critical filling occurs in the vicinity of the bulk critical temperature. In this region we can reasonably expect that universal
properties arise due to the scaling behaviour associated with bulk and surface criticality. In fact, for systems with short-ranged forces there are predictions
for the universal shift of $T_w$ and $T_f$ from $T_c$ which depend only on the half opening angle $\psi=\pi/2-\alpha$ as follows \cite{parry05}:
 \bb
 \frac{T_c-T_w}{T_c-T_f}=R_3(\psi)^{1/\Delta_1}\,,
 \ee
where the universal amplitude $R_d(\psi)$ depends only on the dimension $d$ and the tilt angle and $\Delta_1$ is the surface gap exponent \cite{nakanishi}. At
mean-field level, and for a right-angle wedge the prediction of Landau square-gradient theory is $R_3(\pi/4)=0.518$ (and recall $\Delta_1=1/2$). If we naively
assume that this scaling holds for our model, which recall has long-ranged wall-fluid interactions, this predicts that the filling temperature for the wall
strength $\varepsilon_w=0.8\,\varepsilon$ is $T_f^*=1.364$. This is remarkably close to our numerical value $T_f^*=1.375$ and is indicative that some scaling is
present.


Finally, we mention that a possible change in the order of the filling transition had been predicted by effective Hamiltonian theory \cite{wood2}. However, the mechanism originally proposed for this does not quite apply to the
present DFT model. In the original effective Hamiltonian description, which applied only to rather shallow wedges, the mechanism arose because it was noted that the filling temperature $T_f$ may be below the surface spinodal
temperature $T_{spin}$ at which the activation barrier in the wetting binding potential is first formed. However, this mechanism is only possible if the wall-fluid and fluid-fluid forces have the same range, since it requires
that the Hamaker constant controlling the large distance algebraic decay of $W(\ell)$ changes sign at $T_{spin}$. In the present model, with cut-off LJ fluid-fluid forces and long-ranged wall-fluid forces, no such spinodal
temperature exists and an activation is always present. This is shown explicitly in the inset of Fig.\,\ref{bind08} which shows the binding potential at the filling temperature $T_f$. From this we can conclude that the change
in order is a more general feature of filling transitions that occur close to the bulk critical temperature where the ``short-range'' properties occurring on the scale of the large bulk correlation length can compete with
long-range dispersion forces. This is in keeping with the general expectation that long-ranged forces become less important near the bulk critical point. The observation in our model system that critical wedge filling is
possible even if the walls exhibit first-order wetting, is encouraging that such continuous interfacial transitions can be seen experimentally. This would be particularly interesting because fluctuation effects are far
stronger for critical filling than for critical wetting. For example, for the present case of long-ranged forces (with $p=2$), for which $\ell_w\approx (T_f-T)^{-1/2}$, the interfacial roughness (r.m.s. width) is predicted to
diverge, due to capillary-wave-like fluctuations, with a universal power-law $\xi_{\perp}\approx (T_f-T)^{-1/4}$ which is independent of the range of the forces \cite {wood2}. Of course such fluctuation induced interfacial
roughness is not present in our mean-field DFT and in reality the density profiles $\rho(x,z)$ will be broader near the interface than calculated herein. However this is a minor defect of the mean-field DFT analysis which
should be completely reliable as regards the location of the filling transition, its order and also the determination of the exponent $\beta_w$.

We believe our predictions are testable in the laboratory. For the case of complete wetting there have already been impressive experiments by Mistura and co-workers \cite{mistura} who have verified the leading power law and
amplitude in equation \ref{CoFilling1}. Repeating these experiments with more precisely manufactured wedges would allow one to look at the more subtle next-to-leading order behaviour similar to that described here.
Unfortunately, the materials used so far have precluded the study of fluids which exhibit partial wetting which is of course necessary to see the filling transition. At the moment it appears more likely to us that this
transition can be seen at the micron scale using colloid polymer mixtures similar to studies of wetting and capillary condensation \cite{aarts}.

Our work can be extended in a number of ways. Obviously larger system sizes with finer grids would allow us to probe the critical regime for continuous filling
with greater accuracy. Varying the tilt angle, interaction strengths and range of the forces would also be very informative and would allow us to see whether the
filling transitions for the stronger potentials, where $T_f$ is further from $T_c$, are turned continuous. Generalising our analysis to asymmetric wedges with
competing potentials at each wall would also be straightforward. Finally at low temperatures it would be very interesting to see if one could induce corner
crystalline structure near the wedge apex and defects due to the competition between the lattice directors and the geometrical confinement. We hope that this
work stimulates further 2D and 3D DFT studies of adsorption at structured surfaces and experimental investigations of wedge filling.

\begin{acknowledgments}
 \noindent A.M. acknowledges the support from the Czech Science Foundation, project 13-09914S.
\end{acknowledgments}

\end{document}